\documentclass[12pt]{article}

\usepackage{latexsym}
\usepackage{amssymb}
\usepackage{amsfonts}

\usepackage{hyperref}
\usepackage[dvips]{graphicx,color}
\graphicspath{{images/}}
\usepackage{colortbl}

\def\nc#1{\newcommand{#1}}
\def\rnc#1{\renewcommand{#1}}

\def\a{\alpha}

\nc{\g}{\gamma}
\def\d{\delta}
\nc{\D}{\Delta} 
\nc{\e}{\eta}
\nc{\ep}{\epsilon}

\nc{\ve}{\varepsilon}
\nc{\G}{\Gamma}
\def\k{\kappa}
\nc{\la}{\lambda}
\nc{\La}{\Lambda}
\nc{\om}{\omega}
\nc{\Om}{\Omega}
\nc{\vphi}{\varphi}
\nc{\si}{\sigma}
\nc{\Si}{\Sigma}
\rnc\th{\theta}
\nc\Th{\Theta}
\nc{\z}{\zeta}

\nc{\got}[1]{\mathfrak{#1}} 

\nc\im{{\rm Im}\, }
\nc\re{{\rm Re}\, }

\nc{\Rt}{{\tilde R}}
\nc{\CC}{{\mathbb C}}
\nc\II{{\mathbb I}} 
\nc{\RR}{{\mathbb R}}
\nc{\HH}{{\mathbb H}}
\nc{\NN}{{\mathbb N}}
\nc{\ZZ}{{\mathbb Z}}
\nc{\MM}{{\mathbb M}}

\nc{\ov}[1]{\overline{#1}}

\def\dag{\dagger}

\nc{\non}{\nonumber\\}
\def\nn{\nonumber}
\nc{\noi}{\noindent}

\nc{\p}{\partial}
\nc{\na}{\nabla}
\def\x{\times}

\nc\vev[1]{\ensuremath{\lan #1\ran} {}}
\nc\refeq[1]{(\ref{#1})}
\nc{\eqref}[1]{(\ref{#1})}
\rnc\to[1][]{\ensuremath{\stackrel{#1}{\rightarrow\;}}}

\nc{\twovec}[2]{\left( \!\!
\begin{array}{c} #1\\  #2 \end{array}\!\!\right)}
\nc{\twomat}[4]{\left(\!\! \begin{array}{cc} #1&#2\\ 
#3&#4\end{array}\!\! \right)}

\nc{\ds}{\displaystyle}

\nc{\lan}{\langle}
\nc{\ran}{\rangle}

\nc{\beq}{\begin{equation}}
\nc{\eeq}{\end{equation}}
\nc{\beqa}{\begin{eqnarray}}
\nc{\eeqa}{\end{eqnarray}}
\nc{\beqas}{\begin{eqnarray*}}
\nc{\eeqas}{\end{eqnarray*}}
\nc{\barr}{\begin{array}}
\nc{\earr}{\end{array}}
\nc{\ben}{\begin{enumerate}}
\nc{\een}{\end{enumerate}}
\nc{\bit}{\begin{itemize}}
\nc{\eit}{\end{itemize}}

\nc{\cred}{\color{red}}
\nc{\cblue}{\color{blue}}

\nc\rQ[1][]{\ensuremath{{\cred\leftarrow(?)\mbox{[\footnotesize #1]}} } }

\nc\more{{ \cred{MORE}}}
\nc{\remark}[1]{\cblue[Rem]\footnote[*]{\color{blue}{Remark:} #1}}
\nc{\foot}[1]{{}{\cblue[{}\footnote[0]{\tt #1}]}}
\nc{\ADD}{{\cblue ADD}}

\nc\mfb[1][]{\ensuremath{{\bar 5}_M^{\, #1}}{} }
\nc\yb[2][]{\ensuremath{Y^{#1}_{\overline{#2}}}{}}
\nc\y[2][]{\ensuremath{Y_{#2}^{#1}}{}}
\nc\f[2][]{\ensuremath{5_{(#2)}^{#1}}{}}
\nc\hfb[1][]{\ensuremath{{\bar 5}^{\, #1}_H} }
\nc\et[1]{\ensuremath{10_{(#1)}}{}}
\nc\etb[1]{\ensuremath{\overline{10}_{(#1)}}{}}
\nc\ef[1]{\ensuremath{5_{(#1)}}{}}
\nc\efb[1]{\ensuremath{\overline{5}_{(#1)}}{}}
\nc\ed[2][]{\ensuremath{D^{\, #1}_{(#2)}}{}}
\nc\edb[2][]{\ensuremath{\overline{D}^{\, #1}_{(#2)}}{}}
\nc\yu[2]{\ensuremath{y_{#2}^{(#1)}}{}}

\definecolor{lightgray}{cmyk}{0.1,0.2,0,0.1}

\nc\lgut{\La_{GUT}}
\nc\mpl{M_{Pl}}
\nc\fdx{r_{D/X}}
\def\Ub{\ov U\,}
\def\Db{\ov D\,}
\def\Eb{\ov E\,}

\begin{document}

\title{F-theory inspired GUTs with extra charged matter} 
\author{ J. Pawe{\l}czyk\\{\small {}}\\{\small {\it Institute of Theoretical Physics, University of Warsaw,}}\\{\small {\it ul.\ Ho\.za 69, 00-681 Warsaw, Poland}}\\}
\date{}
\maketitle
\abstract{
\small We consider GUT models inspired by recent local F-theory 
constructions. We show that
 after switching on vevs to scalars the extra matter becomes messengers. We discuss conditions on these vevs under which
the models do not lead to unacceptable
baryon/lepton  number violating processes.}

\newpage
\section{Introduction}

GUT models rise hope for better unification for long time \cite{Langacker:1980}. The basic arguments supporting the idea are twofold: all known matter is organized in SU(5)  multiplets and coupling constants seem to unify at some scale.
It  appears that SUSY GUTs  provide  better coupling unification and shift the unification scale  $\La_{GUT}$ to values acceptable for the proton stability  under heavy gauge bosons exchange \cite{Kounnas:1983,Nanopoulos:1983}. Besides these successes there still remains troublesome problems such as the origin  of the doublet-triplet splitting i.e. phenomenon of absence of a full GUT representation for Higgses of MSSM and the suppression mechanism of the 
baryon/lepton (B/L) number violating processes \cite{Nath:2006}. Strong suppression of these processes is a one of the crucial test for the candidate GUT. For the elimination of dangerous dimension four  operators it is enough to impose e.g. R-parity - an extra symmetry of unknown origin. Suppression of higher--dimensional operates requires extra structure e.g. more symmetries \cite{Nilles:2010}. 

In models constructed recently within the realm of F-theory unification \cite{F-unify}
all known particle physics (excluding gravity effects)
come from a single $E_8$ F-theory singularity
\cite{HV-E8}. The models realize $SU(5)$ GUTs with some extra
 global U(1)'s originating from $SU(5)_\perp/\G$ where $\G$ is  so-called monodromy group. 
 They include SUSY breaking sector and its mediation through
  gauge forces (GMSB), the doublet-triplet splitting is achieved in a novel 
way by introduction of  background fluxes on matter curves of the compact CY space. For some $\G$'s the R-parity is a subgroup of the global U(1)'s. Moreover the global symmetries forbid  dimension five B/L breaking operators. In realistic models these symmetries must be spontaneously broken because messengers masses are provided by vevs of certain scalars. In consequence this may lead  at low energies to generation of dangerous B/L effective operators.
  Besides most of the models of \cite{HV-E8} contain an  extra charged matter which role is unclear at first sight.
  
The purpose of this letter is to discuss all Dirac scenarios F-theory GUTs from \cite{HV-E8}. We shall show that switching on vevs for charged scalars the extra charged matter can be interpreted as messengers and under certain conditions on these vevs  B/L breaking operators generated at low energies are strongly suppressed. All model possesses a bunch of extra neutral scalars  which safely can be assumed to be very massive and decouple.

\newpage
\section{The $\ZZ_3/{\mathbb S}_3$ models}
We start with short description of the model $\ZZ_3$ 
(${\mathbb S}_3$ model is just simple reduction of the latter). Details are in \cite{HV-E8}. The matter content is summarized in the presented table. We must recall that F-theory case the effective Lagrangian contains all the invariant coupling including Yukawas and trilinear terms in Kahler potential (divided by the GUT scale denoted here by $\La_{GUT}$).

\rnc{\arraystretch}{1.2}
\beqa\label{model1}
\begin{tabular}
[c]{|c|c|c|c|c|c|c|c|}
\hline
&\multicolumn{7}{|c|}{Minimal}\\
\cline{2-8}
& $10_{M},\,\y{10}$ & $\overline{5}_{M}$& $5_{H}$ & $\overline{5}_{H}$& $Y_{\overline{10}}^a$ &
$X$ & $N$\\\hline
$U(1)_{PQ}$ & $+1$ & $+1$  & $-2$ & $-2$ &$+3$ & $-4$ & $-3$\\\hline
$U(1)_{\chi}$ & $-1$  & $+3$ & $+2$ & $-2$&$+1$  & $0$ & $-5$\\\hline
\end{tabular}
\rnc{\arraystretch}{1.194}
\begin{tabular}[c]{|c|c|c|>{\columncolor{lightgray}\raggedright}c|c|}\hline
\multicolumn{5}{|c|} {Extra}\\\hline
  $10_{(1)}$  &\yb5 &$Y_{5}$   &\ed1&\edb1 \\ \hline
$0$ & $+1$ & $+3$  &$+1$&$-1$ \\\hline
$+4$& $+3$ & $-3$ &$-5$&$+5$ \\\hline
\end{tabular}
\nn
\eeqa

It must be stressed that the chirality of the  spectrum of 
the model has origin in non--trivial F-theory fluxes through
 2d-cycles where the matter is localized. 
 Manipulating fluxes results in different matter content.
 We shall use this freedom in the paper.
 Existence of such fluxes and cycles is a global issue which
  has not been resolved yet.
 Following \cite{HV-E8} we shall assume that the  appropriate 
 global construction exists.

The model has two extra global U(1) symmetries which are in fact remnants of the "anomalous" gauge symmetries.
They provide selection rules for possible GUT invariants.
It is easy to see that R-parity is subgroup of above: $R=(-1)^q$, where $q=Q_{PQ}$ or $q=Q_\chi$.
We shall slightly modify the matter content compared to the original paper in order to cure SU(5) anomaly.
 The simplest modification is just addition of one extra 
$\yb{10}$ (both fields will be denoted by \yb[a]{10} ,  $a=1,2$).

The model contains standard matter $10_M,\mfb$ as well as appropriate Higgses $5_H,\hfb$. These couple to matter in the conventional way
\beq\label{matt-coup}
W\supset 10_M^25_H,\ 10_M \mfb \hfb
\eeq
Recall that color triplets of Higgs fields get mass thought appropriate hypercharge background flux. Their masses are
 assumed to be of the order $\La_{GUT}$. This will be discussed
  later. Thus  5-dimensional representations of Higgses split 	into
 light doublets and heavy triplets. We shall use somehow hybrid notation
$5_H=(5_H)_2+3_H,\ \hfb= (\hfb)_2+\bar{3}_H$. Of course $(5_H)_2=H_u,\ (\hfb)_2=H_d$ of MSSM.

There is a scalar $X$ which receives non--zero vev $X=\vev X+\th^2\vev{F_X}$ and breaks SUSY. For the discussion of the potential for scalars including $X$ see App.\ref{app:pot}.
Trilinear couplings
\beqa
W&\supset & f_a\yb[a]{10}Y_{10}X,\ \yb5 Y_5 X
\eeqa
through \vev X provides masses for messengers $Y_{10}, Y_5,\yb5$
and one linear combination $f_a \yb[a]{10}$ hereafter called $\yb{10}$ .
The  Kahler potential term $X^\dag 5_H\hfb/\La_{GUT}$ produces $\mu$-term  $\mu=\vev{F_X}/\La_{GUT}$.

What about the extra charged \et1 ? It appears that the model has the following coupling
\beq\label{Yt-et}
W\supset g_a\yb[a]{10}\et1 N 
\eeq
We decompose $g_a\yb[a]{10}$ into the \yb{10} and a new field
\etb1 i.e. $g_a\yb[a]{10}=(\yb{10},\etb1)$ thus
\refeq{Yt-et} provides mass of the order \vev N to the pair  $\{ \etb1,\, \et1\}$  thus turning the fields into extra messengers plus some mixing of the order $N/D$ between \y{10} and \et{10}.

Thus it seems that turning on vev of scalars we just obtain standard GUT model with --minimal messenger sector. One must be careful though. It is apparent that switching on vevs for $X$ and $N$ scalars breaks R-parity what in consequence may lead to baryon/lepton violating processes. It is clear that the smaller are these  vevs the smaller amount of violation one could expect. On the other hand the vevs provide masses for messengers thus there are natural lower bound for their values.
Because we are going to work with an effective action below
 the GUT scale $\La_{GUT}$ we assume that all vevs are much smaller than this scale.
  Thus we introduce small parameters:
$x=\vev X/\La_{GUT},\ n=\vev N/\La_{GUT},\ d=\vev{\ed1}/\La_{GUT}$ (or $d=\vev{\edb1}/\La_{GUT}$ in the second version of the model).

The possible form of the potential for the scalars and it properties including minima and masses are discussed in App. \ref{app:pot}.

\subsection{B/L violation}\label{sec:B-L}
In the rest of the paper we shall discuss effects of switching on vevs of the charged scalars $X,N,D$ i.e. vevs of both bosonic components of the chiral superfields
e.g. for $X:\ \vev X+\th^2\vev{F_X}$, etc.
 This will break both U(1) symmetries spontaneously thus also the R-parity. In consequence it may lead to dangerous processes violating lepton/baryon numbers.
We are going to discuss these issues in the following
section. 

The primary result of non--trivial vevs of $X,N,D$ is mixing between fields of different $Q_{PQ},\ Q_\chi$ charges. The mixing may directly lead to B/L violation.
Let us write down all trilinear coupling between fields charges under GUT group:\footnote{In order not to proliferate coefficients  we denote as $(A,B)$ any liner combination of the fields $ A,\,B$ with coefficients of the order  1. Below we have suppressed obvious conjugate expressions.}\\
$10\x10\x5$ singlets:
\beqa\label{ttf}
(10_M,Y_{10})^2 5_H,\ (10_M,Y_{10})\et1 (\mfb,\yb5)^\dagger\,
\ (10_M,Y_{10})\yb[a\dag]{10}\hfb[\dag],\ \et1\yb[\dag]{10}\y5
\eeqa
$10\x\bar 5\x\bar5$ singlets:
\beq\label{tff}
(10_M,Y_{10})(\mfb,\yb5)\hfb,\ (10_M,Y_{10})\y[\dag]5 5_H^\dagger,\ \yb[a\dag]{10}(\mfb,\yb5)5_H^\dagger,\ \et1\hfb 5_H^\dagger
\eeq
Scanning the above one sees that mixing of \hfb with \mfb and $\et1$ with $10_M$ would lead to B/L violation linear in vev of scalars through $10_M\mfb[2]$ vertex of the superpotential $W$ or $10_M^2\mfb[\dag]$ vertex of Kahler potential the latter being suppressed by the scale $\La_{GUT}$.

Short inspection of the model with \ed1 reveals existence of the following term 
\beq\label{mhd}
W\supset\mfb 5_H \ed1
\eeq
After Higgs triplets are decoupled (see the next paragraph) and $X,\ \ed1$ receive vevs we obtain\footnote{$(\mfb)_2$ denotes doublet of SU(2)	 inside	\mfb and similar for \hfb.}
\beq
W\supset(\mu \hfb+\vev{\ed1}\mfb)_2(5_H)_2 
\eeq
with $\mu=F_X/\La_{GUT}$.
This can be put into canonical form $\mu' \hfb 5_H$ ($\mu'{}^2=\mu^2+\vev{\ed1}^2$) by a rotation: $\hfb\to (\mu \hfb-\vev{\ed1}\mfb)/\mu'$. In consequence
 \refeq{matt-coup} produces  lepton/baryon number  violation vertex
\beq\label{mmm}
y{\vev{\ed1}\over \mu'}\, 10_M \mfb (\mfb)_2
\eeq
where 
we have restored the Yukawa coupling $y$ and the subscript 2 means that we keep only the MSSM doublet piece. The  r.h.s. of the above contains  R-parity breaking operators $\Eb LL,\ QL\Db$ (but not $\Ub\Db\Db$) which couplings are sometimes named $\la,\ \la'$
\cite{Dreiner:1997}. 
The current limits on $\la$'s taken from  \cite{Dreiner:2010}
imply that acceptable values of \vev{\ed1}/$\mu$ are smaller than $10^{-6}$.  But the analysis of our potential for the scalars shows that generically $\vev N\sim \vev{\ed1}$. If we recall that \vev N sets the mass of messengers we immediately conclude that the model is in conflict with phenomenology.\\

The model can be easily cured assuming that the fluxes
 through matter curves are such that the
  spectrum contains \edb1 with opposite U(1) charges
  ($Q_{PQ}=-1,\, Q_\chi=+5$).
If so then instead of \refeq{mhd} we have
\beq\label{mhdb}
K\supset\frac1\La_{GUT}\mfb 5_H \edb[\dag]1
\eeq
This produces mixing with Higgs proportional to F-term of the  superfield \edb1 (which we will denote by ${F_D}$). Hence  \vev{\ed1} is replaced by $\mu_D\equiv {{F_D}/ \La_{GUT}}$ so it is enough that $\mu_D\ll 10^{-6} \mu$
to be in accord with phenomenology. Recalling that $\mu={F_X}/\La_{GUT}$ we obtain
\beq\label{fd-fx}
F_D< 10^{-6} F_X
\eeq
what is reasonable requirement.
The rotation of the Higgs due to \refeq{mhd} is
\beq\label{mix:h2-m2}
(\hfb)_2\to (\hfb)_2-{{F_D}\over {F_X}}\,(\mfb)_2
\eeq
From now one we are going to focus on this version of the model \refeq{model1}.\\ 

At this point let us discuss at some length the influence of Higgs color triplets. Their mass term is
\beq
M 3_H\tilde{3}_H+M'\ov 3_{H}\tilde{\ov 3}_{ H}+\mu\, 3_H\ov 3_H+ X \tilde{3}_H\tilde{\ov 3}_{ H}
\eeq
where tilde fields are appropriate KK modes of F-theory compactification. Due to their charge they may   couple to $X$ too.
Of course the latter will obtain vev: $X\to \vev X$. $M$ and $M'$ are masses of the order $\La_{GUT}$. Adding the mixing \refeq{mhdb} and diagonalizing the mass term we get
rotation
\beq\label{mix:h3-m3}
\ov 3_H\to \ov 3_H-\frac{\mu_D\,X}{M^2}(\mfb)_3
\eeq
and the effective B violating vertices
\footnote{Below we use the standard notation for MSSM models where $\Db$ denotes chiral superfield containing the down quark. I hope the reader will not confuse it with the scalar \edb{1}. }
\beq
-\frac{\mu_D\,X}{M^2} (\Ub\Ub\Db, QL\Db).
\eeq
With $\vev X\sim 10^{-2}\La_{GUT},\ \mu_D\ll 10^{-6} \mu\sim 10^{-20}\La_{GUT},\ M\sim \La_{GUT}$ the suppression factor $\ll 10^{-22}$ is in agreement with phenomenology (see also \cite{Nath:2006} Table 2.).

This ends discussion of dimension 4 B/L--violating vertices which may appear in the model discussed.

\subsection{Higher--dimensional operators}

Here we are going to look for possible higher dimension B/L breaking operators.  There is one dangerous dimension 5 operator in superpotential invariant under SU(5): $10_M^3\mfb$ which includes two dangerous MSSM operators: $ QQQL,\ \Ub\Ub\Eb\Db$.
The operator is invariant under $U(1)_\chi$ but not under $U(1)_{PQ}$.  The possible dimension 6 operators are numerous and they may correct the superpotential as well as the Kahler potential.
%
The primary source of these operators are exchange of heavy states with appropriate group structure. The universal contribution comes form heavy GUT gauge fields and their KK modes - these where discusses in e.g. \cite{H-rev} and they contribute to the Kahler potential only. 
The exchange of heavy color Higgses is strongly suppressed: the reasoning goes in similar way as presented in the previous section (see also \cite{H-rev}). The Higgs doublets gives no effects.

The remaining possibility are diagrams with exchange of messengers. Here we shall show that dimension 5 operator is not produces and that the only nonvanishing contribution is a dimension 6 correction to the Kahler potential with a very small coefficient.

Hence one has to find out all operators of the form
$M M' Y$ where $M$'s denote matter fields and $Y$ a messenger.
These operators appear
 as a result of mixing discussed in the previous section.
We shall be interested in operators arising from {\em single} redefinition because each redefinition is accompanied by small factor 
of the order 
$n\equiv \vev N/\La_{GUT},\ d\equiv \vev{\edb1}/\La_{GUT}$ or $ {F_D}/{F_X}$(see above).

Let us discuss the remaining (besides \refeq{mix:h2-m2}) mixings between fields. The couplings of interest are\footnote{$\yb5 5_H \edb[\dag]1$ has negligible effect.}
\beqa\label{mass-5}
W&\supset& \quad \hfb \y5 \vev{\edb1},\ \mu\, \hfb 5_H,\ \vev X \y5\yb5,\ \mu_D \yb5 5_H\\
K&\supset & (k1):\ {N\over \La_{GUT}}\,(\mfb,\yb5)\hfb[\dag],\
(k2):\ {\edb1\over \La_{GUT}}\, (10_M,Y_{10})^+ \et1.
\eeqa
where $\mu=F_X/\La_{GUT},\ \mu_D=F_D/\La_{GUT}$.
Redefining 	
$
\hfb\to\hfb-{N\over \La_{GUT}}(\mfb,\yb5) 
$
one can get rid of (k1)\footnote{We ignored here subleading	terms from Eqs.(\ref{mix:h2-m2},\ref{mix:h3-m3}).} in the expense of $-|{N\over \La_{GUT}}\,(\mfb,\yb5) |^2$. 
The latter can be completely removed  when $N\to \vev N$ (what is assumed hereafter) redefining  kinetic terms  for \mfb,\yb5. 
This results in $\mfb\to \mfb+|n|^2 \yb5$ and small rescaling  of $\yb5$ both irrelevant for our analysis. Furthermore the rotation: 
$
\hfb\to\hfb+n \mfb,\ \mfb\to\mfb-n \hfb
$
adds up to: $\hfb\to\hfb-n \yb5,\ \mfb\to\mfb-n \hfb$.
Next we diagonalize the mass terms \refeq{mass-5} which lead to irrelevant mixing between Higgses and messengers.
Hence we are going to ignore these contributions.
Similarly treatment  of (k2) gives
\beq\label{re-et1}
\et1\to \et1- d\,\y{10},\ 10_M\to 10_M-d\, \et1
\eeq
Scrutinizing \refeq{ttf} one finds the following MM'Y couplings
\beqa
\k_1 10_M\et1\mfb[\dag],\quad
\k_2 10_M^\dagger \yb[a]{10}(\mfb)_2
\eeqa
where $
\k_1=1/\La_{GUT},\, 
\k_2=F_D/(F_X\La_{GUT})$
while from \refeq{tff} one obtains
\beq
\yu1{10}\y{10} \mfb(\mfb)_2,\ \yu15 10_M \mfb \yb5.
\eeq
where $\yu1{10}=F_D/F_X,\ \yu15=max(F_D/F_X,n)$. 
Integrating over the messengers $\y{10}$ we obtain only a single dimension 6 operator. Suppressing the family indices it has the form
\beq
\d K\supset y_1y_2({(F_D/F_X)^2\mpl\over \La_{GUT}^3})
10_M^\dagger \mfb\mfb(\mfb)_2+c.c.
\eeq
where $y_i$ are the i-th family Yukawas.
The above contains such B/L breaking MSSM operators as 
$Q^\dag \Db^2L$. 
The effective coupling constant 
is of the order $y_1y_2 10^{-8}/\La_{GUT}^2$. 
Taking into account that the Yukawa couplings  
for the first family can be  as small as $y_1\sim 10^{-5}$ 
we obtain enormous suppression.
\section{$\ZZ_2/\ZZ_2\x \ZZ_2$ model}	

Here we are going to discuss the $\ZZ_2$ model of \cite{HV-E8}. The $\ZZ_2\x \ZZ_2$ model is a reduced version thereof
thus our analysis will work also in this case.  The minimal matter is the same as in $\ZZ_3/{\mathbb S}_3$ model of the previous section and it will not be displayed here. 
The  possible extra  matter is presented in the table.
We are going to shorten the discussion  here 
to issues related to that extra matter. 

 We shall denote the minimal \hfb as \hfb[1] and \efb1  as \hfb[2].
\beq
\begin{tabular}
[c]{|c|c|c|>{\columncolor{lightgray}\raggedright}c|c|}\hline
Extra & $10_{(1)}$ &\hfb[2] & \efb2 &
$\ef3$\\\hline
$U(1)_{PQ}$ & $+4$ & $-2$ &$+5$ & $+6$\\\hline
$U(1)_{\chi}$ & $+4$& $-2$  & $+3$ & $+2$\\\hline
\end{tabular}
\begin{tabular}
[c]{|>{\columncolor{lightgray}\raggedright}c|c|}\hline
 \ed2  & $D_{(4)}$ \\\hline
 $+4$ & $-7$ \\\hline
$0$ & $-5$ \\\hline
\end{tabular}
\eeq
The apparent differences lie in the
 distribution of charges among 
 the extra matter \et1 and in the sector of 5's. 
 
 First one must notice that there is no way one can give mass to
 \efb2 without serious distraction done for the minimal sector.
 Thus we assume the  field is absent from the spectrum.
 Similarly we remove cumbersome \ed2 which could form a mass 
 term with $X$.
 We guess the extra pair of 5's will become messengers.
 The relevant couplings producing mass term 
are
\beq
W\supset  {5}_{(3)} (\mfb \ed4+f_a\hfb[a] X)
\eeq
With obvious definition of $\a$  the
messenger is
($\yb[(3)]5\sim\cos\a\, (f_a\hfb[a])+\sin\a\,\, \mfb$).
The light $\ov 5$ 's are $\mfb[\,']\sim\cos\a\,\mfb-\sin\a 
\, (f_a\hfb[a])$ and $\hfb[\,']\sim\ep_{ab}f^a\hfb[b]$.
We expect that $\cos\a\approx 1$ i.e. the matter \mfb field will not vary much during the process of redefinition. To define physical Higgs and matter $\ov 5$ 's we need take into account  the only trilinear coupling in the Kahler potential
\beq
5_H(\hfb[1],\hfb[2])X^\dag\sim 5_H(-\sin\a\,\mfb[\,'], \hfb[\,'])X^\dag
\eeq
This finally defines MSSM Higgs \hfb[f]$=(-\sin\a\,\mfb[\,'], \hfb[\,'])\approx (-\sin\a\,\mfb[f], \hfb[\,'])$.

As in the previous section the mixing between fields may generate dangerous B/L--violating vertices. Below we estimate the coupling constant of the leading dimension 4 operator.  
The operator of interest originates from $(\hfb[1],\hfb[2])10_M\mfb $
producing: 
$
\sin\a\ \mfb[f] 10_M\mfb[f]
$
what gives
\beq
{\vev{\ed4}\over \vev X}\ll 10^{-6}
\eeq
We expect that the analysis of higher--dimensional operators will give negligible B/L--violating effects.

It is easy to see that the new \et1 is  the messenger coupled to
\yb[a]{10} and \ed4 thus receiving mass when the scalar \ed4 acquire a vev.

\section{Conclusions}
The discussion presented shows that the F-theory GUT models of   \cite{HV-E8} seem to by phenomenologically viable after small
 (but sensible from the point of view of F-theory) 
  modifications i.e.  at low energies they give MSSM with some
 extra sterile scalars and broken SUSY. Apparent lack of
 R-parity spontaneously broken just below the GUT scale does
not lead to dangerous B/L breaking processes under some
 conditions put on scalar vevs. Of course the simple
 analysis presented in this paper does not say anything
 about such important issue as FCNC, dark matter
  candidates, soft-SUSY breaking terms and more. This
   would require deeper studies which go beyond this letter.

\newpage
\appendix
\section{Appendix: scalar potential}\label{app:pot}
Here we shall discuss the potential for the scalars leading to SUSY breaking \cite{sspot,Nomura,Kitano}.	 
We focus on $\ZZ_3$ model as the discussion for  the ${\ZZ}_2$ model would be  very similar.
According to the results of Sec.\ref{sec:B-L} we must work with the version of \refeq{model1} with \edb1 field.
Also it is necessary  that in a global setting  there will be instantons generating Polonyi  terms for all the scalars \cite{inst}.
 Gauge invariance forces the Polonyi terms to be accompany by appropriate closed string modes (denoted here after by $t$) which we choose here to be twisted moduli \cite{JLP}. 
One could consider untwisted moduli too but then in order to achieve viable vacua one needs to generate FI-term \cite{Komargodski:2009pc} as in \cite{HVS}. We shall not work out this possibility because this section serves merely as the illustration of the SUSY breaking generation mechanism. 
\beq
W=W_0+f_X\, e^{-t_{PQ}}X+f_N\, e^{-\frac34 t_{PQ-t_\chi}}N
+f_D\, e^{-\frac14 t_{PQ}+t_\chi}\edb1
\eeq
The Kahler potential except the standard piece $K_0$ gets contribution from the trilinear coupling $X^+\edb1 N$ as well as corrections due to the exchange of the  anomalous U(1) gauge bosons.
\beqa
K&=&K_0+\frac1\La_{GUT} (X\edb[\dag]1 N^++ X^+\edb1 N)\non
&&-\frac{g^2}{4\La_{GUT}^2}((|X|^2+\frac34|N|^2+\frac14|\edb1|^2)^2+(|N|^2-|\edb1|^2)^2)
\eeqa
Finally there are D-terms
\beqa
D_{PQ}&=&|X|^2+\frac34|N|^2+\frac14|\edb1|^2+
\la^2(t_{PQ}+\bar t_{PQ})\\
D_{\chi}&=&|N|^2-|\edb1|^2+
\la^2(t_\chi+\bar t_\chi)
\eeqa
where $\la$ is the mass scale characterizing the anomalous massive gauge bosons. We expect $\la$ to be close to $\La_{GUT}$. With $l\equiv\La_{GUT}/\la$ the extreme of the potential for the scalars are
\beqa
\vev X&=&   \frac{2 }{g^2 +l^2} \,{w_0}\La_{GUT}^2 \non
\vev N&=& \a_1\,{\frac{f_D}{f_X}} \La_{GUT}+\a_2\,{\frac{f_N}{f_X}} \, w_0\La_{GUT}^2
\\\label{vevs}
\vev{\edb1}&=& \a_3\,{\frac{f_N}{f_X}} \La_{GUT}+\a_4\, {\frac{f_D}{f_X}}\, w_0 \La_{GUT}^2
\nn
\eeqa
where $w_0=W_0/f_X$ and coefficients $\a_i$ are of the order one.\footnote{ Explicitly
\beqa
\frac{8 }{3 g^2+6 l^2+8},\ \frac{2 \left(72 g^2+5 g^4+96 l^2+18 g^2 l^2+16 l^4\right)}{\left(g^2+l^2\right) \left(g^2+2 l^2+8\right) \left(3 g^2+6 l^2+8\right)}\non
\frac{8 }{g^2+2 l^2+8},\ \frac{2 \left(88 g^2+21 g^4+96 l^2+66 g^2 l^2+48 l^4\right)}{\left(g^2+l^2\right) \left(g^2+2 l^2+8\right) \left(3 g^2+6 l^2+8\right)}
\nn
\eeqa
}
Consistency of the calculations require that $\vev X,\, 
\vev N,\, \vev{\edb1}\ll\La_{GUT}$ thus we need $f_N,\, f_D\ll f_X$.
Notice that generically  vevs of $N,\ \edb1$ are related: none of them vanish without fine tuning.
The phenomenological constraint \refeq{fd-fx}
$F_D< 10^{-6} F_X$ implies $f_D < 10^{-6} f_X$ thus also $\vev N\ll \vev X$. The value of
$f_N$ is unconstraint thus also \vev{\edb1}. 
The contributions to gauginos masses are  $\frac{g^2}{16\pi^2} f_X/\vev X$ and 
 $\frac{g^2}{16\pi^2} f_N/\vev N$. Due to smallness of $f_N/f_X$ we can neglect it in \refeq{vevs} obtaining $f_N/\vev N\sim f_X/\La_{GUT}\sim f_X/\vev X$  thus enhancing the GMSB mechanism.
All scalars have similar masses for $f_X\gg f_N$
\beq
m_{X}^2=g^2\frac{f_X^2}{\La_{GUT}^2},\ m_{N}^2=(8+3g^2)\frac{f_X^2}{8\La_{GUT}^2},\ 
m_{D}^2=(8+g^2)\frac{f_X^2}{8\La_{GUT}^2}
\eeq
With $f_X\sim 10^{-18} \mpl^2,\ g\sim 0.3$ and $\La_{GUT}=10^{-2} \mpl$ one gets
$m_{N,D}\sim 100$ GeV, $m_X\sim 30$ GeV.

\newpage
\section*{Acknowledgments}
\vspace*{.5cm}
\noindent The author would like to acknowledge  stimulating discussions with Emilian Dudas, Jonathan Heckman, Tomasz Jeli\'nski, Pran Nath, Stefan Pokorski, Tomasz Taylor and 
Krzysztof Turzy\'nski.  
\noindent This work was partially supported by the 
EC 6th Framework Programme MRTN-CT-2006-035863 , TOK Project  
MTKD-CT-2005-029466 and   Polish Ministry of Science MNiSW grant
under contract N N202 091839 (2010-2013).

\end{document}